\documentclass[preprint,3p,11pt,onecolumn]{elsarticle}
\usepackage{natbib} 
\usepackage{amssymb}
\usepackage{amsmath} 
\usepackage{caption}
\usepackage{graphicx}
\usepackage{xcolor}

\usepackage[colorlinks,allcolors=blue]{hyperref}

\usepackage{subcaption}
\usepackage{adjustbox}
\usepackage{svg}

\definecolor{bleudefrance}{rgb}{0.19, 0.55, 0.91}

\usepackage{color}

\journal{Additive Manufacturing}

\usepackage{siunitx}

\begin{document}

\begin{frontmatter}



\title{Combined thermal and particle shape effects on powder spreading in additive manufacturing via discrete element simulations}

\author{Sudeshna Roy\footnote{Equal contribution.}$^{1*}$}
\author{Hongyi Xiao\footnotemark[1]$^{1,2}$}
\author{Vasileios Angelidakis\footnotemark[1]$^{1,3}$}
\author{Thorsten P\"oschel$^{1}$}
\address{
1. Institute for Multiscale Simulation, Friedrich-Alexander-Universit\"at Erlangen-N\"urnberg, Erlangen, Germany
\\ 2. Department of Mechanical Engineering, University of Michigan, Ann Arbor, USA
\\3. School of Natural and Built Environment, Queen's University Belfast, Belfast, UK}
\cortext[correspondingauthor]{Correspondence: sudeshna.roy@fau.de (S.R.)}

\begin{abstract}
The thermal and mechanical behaviors of powders are important for various additive manufacturing technologies. For powder bed fusion, capturing the temperature profile and the packing structure of the powders prior to melting is challenging due to both the various pathways of heat transfer and the complicated properties of powder system. Furthermore, these two effects can be coupled due to the temperature dependence of particle properties. This study addresses this challenge using a discrete element model that simulates non-spherical particles with thermal properties in powder spreading. Thermal conduction and radiation are introduced to a multi-sphere particle formulation for capturing the heat transfer among irregular-shaped powders, which have temperature-dependent elastic properties. The model is utilized to simulate the spreading of pre-heated PA12 powder through a hot substrate representing the part under manufacturing. Differences in the temperature profiles were found in the spreading cases with different particle shapes, spreading speed, and temperature dependence of the elastic moduli. The temperature of particles below the spreading blade is found to be dependent on the kinematics of the heap of particles in front, which eventually is influenced by the temperature-dependent properties of the particles.
\end{abstract}


\end{frontmatter}


\section{Introduction}
Powder-based additive manufacturing is an effective and sustainable manufacturing technique that has the potential to manufacture objects of arbitrary and complex shapes \citep{bhavar2017review,vock2019powders,singh2021powder}. One specific method with promising development is powder bed fusion, where a laser or an electron beam selectively melts sequentially deposited layers of powders to form intricate parts upon solidification, while the un-melted powders act as support. Although this method presents many merits, such as rapid prototyping, low energy consumption and reduced material waste, compared to traditional manufacturing techniques, the manufacturing quality of the built part is far from ideal with the preparation of the powder bed being one of the major sources of problems. 

The powder packing before melting is crucial for the part quality \citep{mostafaei2022defects}, and it is affected by the recoating process \citep{nasato2017numerical,chen2019powder,nasato2021influence}, which typically involves a blade or a roller spreading a layer of powder over a loose powder bed or previously sintered powders. Powders are out-of-equilibrium complex materials often made of cohesive particles with irregular shapes that typically have poor flowability. The packing structure achieved in spreading is highly dependent on the material properties and process parameters \citep{he2020linking,shaheen2021influence}, which are difficult to optimise, as powder flow is far more complicated than that of Newtonian liquids. As a result, problems exist such as voids sustained in the powder bed due to friction, interlocking due to irregular particle shape, or cohesion \citep{roy2023effect,roy2022local}. Although high-speed powder spreading can efficiently enhance the productivity of laser powder bed fusion, it also reduces the packing density of powder layer and dimensional accuracy in the building direction, which is unfavorable for the part quality in additive manufacturing. Consequently, a higher bed temperature is achieved before the next layer melting, reducing fusion defects and minimizing temperature gradients, which is an advantage of high-speed spreading \cite{chen2022high}.
For polymer powders that are amorphous or semi-crystalline, their mechanical properties can have rather strong temperature dependence. This is highly relevant as their temperatures are elevated significantly during spreading due to combined effects of pre-heating, heat lamps, and spreading on printed parts. This effect can further compromise the quality of the powder layer, making obtaining a dense powder bed with smooth surface even more challenging.

Understanding the influence of temperature-dependent properties for irregularly-shaped particles can be challenging at the particle-level, as it is hard to probe in-situ powder particles experimentally, due to their small sizes. Multiphysics simulations that simultaneously model the mechanics and heat transfer can be utilised to study the spreading process, which often treat the powder as a continuum phase with bulk properties. However, it is can be difficult for continuum-based computational models to include these effects, as the relation between microscopic details and continuum-level properties are missing, such as how the bulk stiffness, density, and heat transfer parameters can be deduced from material properties for a packing of complex-shaped particles. To resolve particle-level physics in power spreading, simulations using the Discrete Element Method (DEM) that directly calculate trajectories and interactions of individual particles have proved to be effective. The heat conduction between spherical particles has also been simulated with discrete element methods \citep{peng2020heat}, which considers the conduction (e.g. \citep{batchelor1977thermal,feng2009discrete,zhou2009particle,kiani2019thermal,adepu2021wall}), convection (e.g. \citep{zhou2009particle,zhou2023cfd}), and radiation (e.g. \citep{zhou2023cfd,zhou2009particle,cheng2013particle,wu2016effect}) between the particles and the ambient environment.

Real powder particles feature a variety of irregular shapes. Particle shape affects the packing density and packing structure of granular assemblies, as irregular particles tend to form complex structures compared to spherical particles. Particle shape exhibits various aspects based on the considered level of observation, which typically entail form, roundness and surface texture \citep{blott2008particle,angelidakis2021shape}. Particle form plays a crucial role in the way irregular particles pack into many-particle assemblies, and is typically characterised in terms of elongation, flatness and compactness, which are used to classify particles as flat, elongated, bladed (i.e. both flat and elongated) and compact \citep{Angelidakis2022elongation}. Different particle shapes demonstrate different packing characteristics; for instance, flat particles tend to align their largest projected area to a surface, when deposited under gravity, elongated particles tend to align their longest dimension to the surface, while compact particles have no clear preference based on their form characteristics alone. In the context of powder spreading processes, particle shape influences the surface roughness of the produced powder layer, which in turn affects the part quality of the object manufactured via the powder fusion process \citep{nasato2020influence}. While irregularly shaped particles offer a larger exposed surface area, leading to improved conductivity within the powder bed, they come with inherent disadvantages, most notably reduced flowability. This limitation poses a challenge to the efficiency of additive manufacturing processes, particularly in applications like laser beam melting, where factors such as processing speed and material selection are crucial \cite{goodridge2012laser}. To overcome the issue of reduced flowability, a strategic shift to the utilization of spherical particles with a narrow size distribution becomes advantageous \cite{schmidt2016optimized, blumel2015increasing, ziegelmeier2015experimental, dadbakhsh2016effect, sachs2015rounding, yang2019preparation}. These spherical particles, often derived through innovative production methods such as melt emulsification \cite{fanselow2016production} and wet grinding with a rounding process \cite{schmidt2014novel}, are preferred in additive manufacturing processes due to their ability to enhance flowability and mitigate the drawbacks associated with irregular shapes. Specifically, spherical particles of polyamides also find widespread use, contributing to the overall efficiency of the additive manufacturing process.

This work addresses the combined effects of heat transfer and particle shape within a powder-layer-spreading framework. To this end, a multi-sphere algorithm \citep{ostanin2024rigid} is coupled with a thermal particle formulation in the \texttt{MercuryDPM} \citep{shaheen2023thermal} open-source software package. First, imaging data of polyamide 12 (PA12) particles are used to generate three-dimensional multi-sphere particles of realistic shape, using a recently proposed open-source platform for particle generation. Then, mixtures of these realistic multi-spheres with thermal properties are used to used to simulate a powder spreading process to investigate the effects of varying blade speed and particle stiffness on the powder bed temperature profile, both during the powder spreading phase and in the time immediately after spreading. To explore the effect of particle shape in the thermal properties of the system, powder bed temperature profiles are compared for different particle shapes, namely for the realistic irregular shapes versus spherical particles of same size.
Conclusions are drawn on the effect the varied parameters have on the powder flowability and, consequently, on the temperature profile of the produced powder layers.

\section{Methods}
\subsection{Multi-sphere particle generation}
Several simulation techniques exist to model non-spherical particles in the Discrete Element Method (DEM). The multi-sphere approach is one of the oldest and more versatile techniques, where a collection of spheres are connected rigidly to form irregular particles. In this approach, individual subspheres are only used to facilitate contact detection, while the collection of spheres is considered as one particle when it comes to calculating inertial characteristics and integrating the motion of the particle. The number and size of the spheres used to create a multi-sphere particle defines its morphological fidelity, i.e. the degree in which it approximates the morphology of a target particle shape, as well as the computational cost associated to its interactions with neighbouring particles.

\begin{figure}[hbtp]
\centering
{\adjustbox{trim=0cm 2.5cm 0cm 2cm,clip}
{\includegraphics[width=\textwidth]{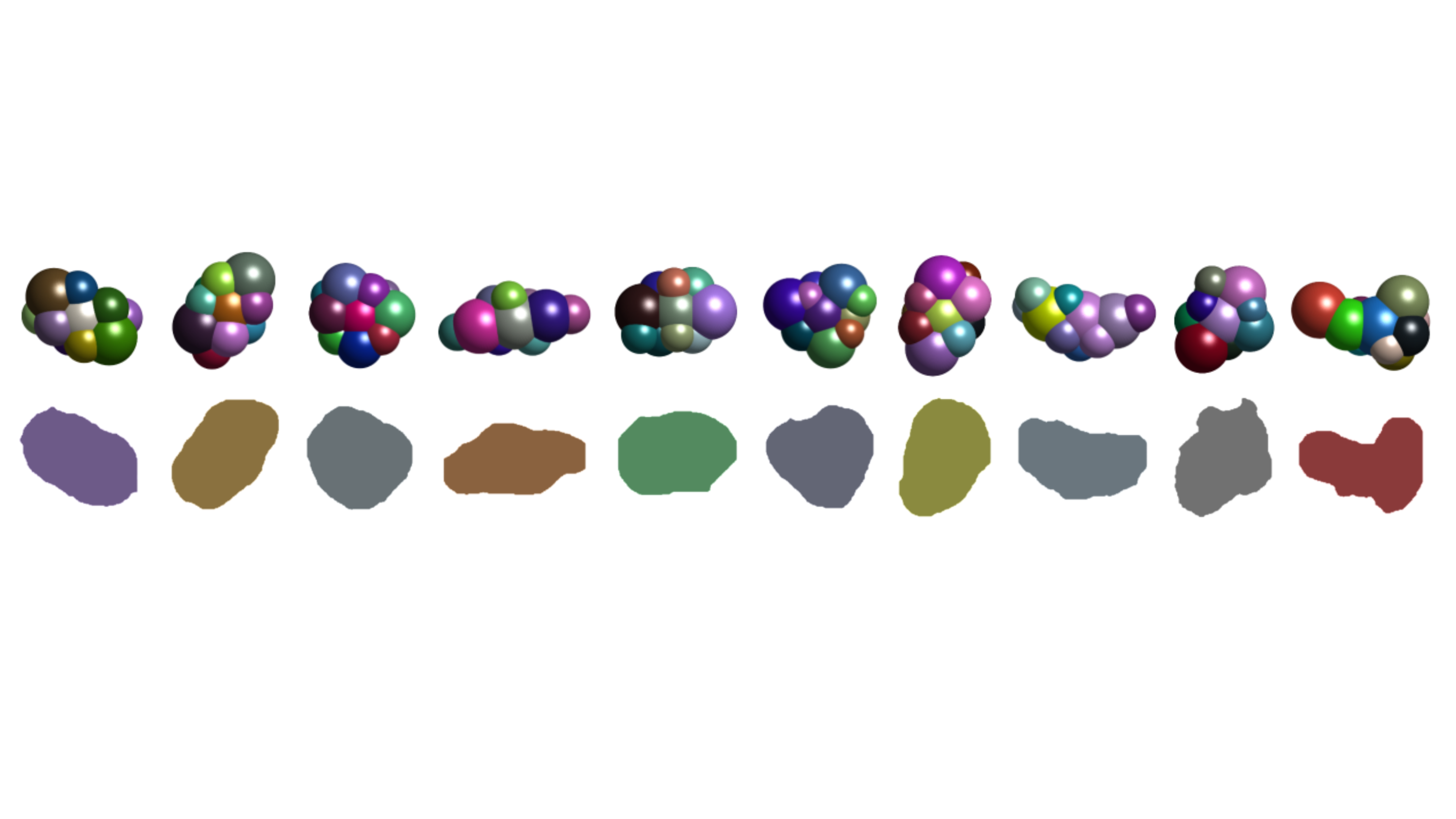}}}
\caption{Non-spherical PA12 powder particles. Multi-sphere particles generated based on binarised SEM images of the material.}
\label{fig:PA12_particles}
\end{figure}

In this work, multi-sphere particles of PA12 are generated to approximate the shape of real powder particles. To this end, an image-informed particle generation procedure is put in place, where scanning electron microscopy (SEM) images of PA12 particles \citep{nasato2021influence} are extruded into three-dimensional ones, which are then used as a target geometry for the generation of realistic multi-sphere particles. Multi-sphere generation is carried out utilising a recently developed particle generation method based on the Euclidean transform of three-dimensional images. In particular, the open-source software \texttt{CLUMP} is employed \citep{angelidakis2021clump}, which provides various techniques for multi-sphere generation. \autoref{fig:PA12_particles} shows ten multi-spheres generated using the SEM imaging data of PA12 powders. The number of subspheres, here $N_\mathrm{sub}=10$, is chosen so that the multi-spheres approximate adequately the particle morphology compared to the imaging data, while its effect on the thermal particle properties is explored in this paper.

\subsection{Thermal contact model}
\begin{table}[h!]
    \centering
    \caption{Material and model parameters.}
    \label{tab:modelling_parameters}
    \begin{tabular}{||c | c | c||} 
    \hline
    Properties & Value & Reference \\ [0.5ex] 
    \hline\hline
    Particle diameter $D_{50}$ & $55 ~\mu$m & \cite{parteli2016particle,nasato2021influence} \\ 
    Density $\rho$ & $1000$ kg/m$^{-3}$ &  \cite{nasato2021influence} \\
    Young's modulus $E$ & $2.3\times{10}^6$ Pa &  \cite{salmoria2012mechanical,parteli2016particle} \\
    Poisson ratio $\nu$ & $0.40$ &  \cite{nasato2021influence} \\
    Dissipation coefficient $A_n$ & $1.85\times{10}^{-7}$ s & \cite{muller2011collision} \\
    Coulomb's friction coefficient $\mu$ & $0.50$ & \cite{nasato2021influence} \\
    Latent heat of fusion $L_f$ & $101.66\times{10}^3$ J/kg& \cite{kulinowski2022development} \\ 
    Solid heat capacity $c_p$ & $1200$ J/{kgK} & \cite{hejmady2019novel} \\ 
    Thermal conductivity coefficient $k_s$ & $0.12$ W/mK & \cite{hejmady2019novel,yuan2011thermal,laumer2014fundamental} \\
    Thermal convection coefficient $h_\mathrm{conv}$ & $15$ W/m$^2$K & \cite{riedlbauer2015thermomechanical,peyre2015experimental,yaagoubi2021review} \\ 
    Emissivity $\epsilon$ & $0.80$ & \cite{,yaagoubi2021review,belliveau2020mid,peyre2015experimental} \\ 
    Melting temperature $T_m$ & $451.15$ K & \cite{hejmady2019novel} \\ 
    Temperature interval $\Delta{T}$ & $20$ K & \cite{ganeriwala2016coupled,muhieddine2009various} \\ 
    Ambient temperature & $393$ K & \cite{hejmady2019novel} \\ 
        Particle temperature & $393$ K & \cite{peyre2015experimental,laumer2015influence,li2021experimental} \\ [1ex] 
    \hline
    \end{tabular}

        

\end{table}

The contact behaviour between the powder particles is described using the simplified, viscoelastic Hertz-Mindlin contact law (no-slip solution), which is enhanced with a temperature dependence. The normal contact force is given by

\begin{equation}
    \vec{F_n} = \min\left(0, -\rho \xi^{3/2} - \frac{3}{2}A_n\rho\sqrt{\xi}\dot{\xi}\right) \vec{e_n}
    \label{Eq:Hertz_normal}
\end{equation}


with

\begin{equation}
    \rho = 4/3 \, E^{*}_{T} \, \sqrt{R^*} 
    \label{Eq:Hertz_rho}
\end{equation}

where $E^*_T$ is the temperature dependent effective Young's modulus and $R^*=R_1R_2/\left(R_1+R_2\right)$ is the effective radius. For contact between two identical materials, \autoref{Eq:Hertz_rho} simplifies to $\rho = \frac{2E\sqrt{R^*}}{3\left(1-\nu^2\right)}$, where $E$ and $\nu$ the Young's modulus and Poisson ratio of the material, respectively. The Young's modulus considered here is two scale of magnitude smaller than the real Young's modulus of PA12 to decrease computational time, as in Parteli and P\"oschel \cite{parteli2016particle}.

The employed viscous damping model is used to simulate non-constant normal restitution for different impact velocities between particles, in alignment with experimental evidence \cite{muller2011collision}. The dissipative constant A is calculated for a given coefficient of restitution at a given impact velocity, based on the particle elastic properties and size, as in \cite{muller2011collision}. Here, a coefficient of restitution of 0.7 is considered at a characteristic impact velocity set to 50 mm/sec, which is the blade speed of the powder spreading process.

A temperature-dependent effective Young's modulus $E^*$ is considered, which varies during the simulation, based on the current temperature of the particles in contact, as in \cite{luding2005discrete,shaheen2023thermal}.

\begin{equation}
    E^{*} =\frac{E^{*}_0}{\alpha} \left[(1+(\alpha-1)\tanh{\left(\frac{T_m-T}{\Delta{}T}\right)}\right]
\label{eq:functionModulus}    
\end{equation}

where $\alpha$ is the degree of Young's modulus change near melting temperature, $E^{*}_{0}=\left(\frac{1-\nu_i^2}{E_i} + \frac{1-\nu_j^2}{E_j}\right)^{-1}$ is the effective Young's modulus at room temperature, $T_m$ is the melting temperature, $T=\frac{T_i+T_j}{2}$ is the average temperature between the two particles in contact and $\Delta{}T$ is the temperature interval around the melting temperature, which is used to calculate the energy required for phase change, as in \cite{shaheen2023thermal}.

In the original implementation of this thermal model by \citep{shaheen2023thermal}, a temperature-dependent Hertzian model was combined with a linear tangential force formulation of constant stiffness, to facilitate the calculation of particle-particle and wall-particle contact forces. Here, we extend their formulation, to implement a fully-temperature-dependent Hertz-Mindlin contact law. The contact force in the tangential direction is given by

\begin{equation}
    \vec{F_t} = -\min \left[ \mu|\vec{F_n}|,  \int_{path}^{} 8 G^{*}_{T}\sqrt{R^* \xi} \,ds 
    + A_t  \sqrt{R^* \xi} v_t \right] \vec{e_t}
\end{equation}

where the the no-slip solution of Mindlin \cite{mindlin1949compliance} is used to calculate the elastic tangential contact forces, and the approach of Parteli and P\"oschel \cite{parteli2016particle} is used to calculate the tangential dissipative constant $A_t \approx 2 A_n E^*$. Regarding the elastic material parameters,  $G^{*}_{T}$ is the temperature-dependent effective shear modulus. For non-thermal particles, the effective shear modulus is calculated as $G^{*}=\left(\frac{2-\nu_i}{G_i} + \frac{2-\nu_j}{G_j}\right)^{-1}$ which for two identical materials simplifies to $G^{*}=\frac{4G}{2-\nu}$. To reflect the temperature dependence of the shear modulus, the following assumption was made in this study: According to the Hertz-Mindlin contact model (no-slip solution), the shear to normal stiffness ratio is defined as $\kappa=k_t/k_n$, where the tangential elastic stiffness for non-sliding contacts is $k_t={dF_t}/{d_s}=8G^{*}\sqrt{R^*\xi}$ and the normal elastic stiffness is $k_n={dF_n}/{d\xi}=2E^{*}\sqrt{R^*\xi}$, thus leading to $\kappa=k_t/k_n=4G^{*}/E^{*}$. Considering the stiffness ratio constant for the PA12 material, the evolving, temperature-dependent shear modulus is calculated as $G^{*}=\kappa E^{*}/4$, which changes throughout the simulation, according to the change of the Young's modulus. For two identical materials in contact, the stiffness ratio can be written as $\kappa=\frac{2\left(1-\nu\right)}{\left(2-\nu\right)}=$ which highlights that the current modelling approach assumes no change of the Poisson ratio during heating of the PA12 particles. A future extension of the model would be to update the stiffness ratio with a potential thermal-dependent evolution of the Poisson ratio, throughout the simulation, if such an effect is established in the literature.

Cohesion between multi-sphere particles are modelled via adding the Johnson-Kendall-Roberts (JKR) model \citep{johnson1971surface} as an attractive force when two slave particles come into contact:

\begin{equation}
    \vec{F}_\mathrm{JKR} = 4\sqrt{\pi a^3\gamma E^*} \; \vec{e_n}
\end{equation}
where $\gamma$ is the surface energy density and $a$ is the contact radius related to deformation, calculated using $\xi = \frac{a^2}{R^*} - \sqrt{\frac{4\pi a\gamma}{E^*}}$. The maximum interaction distance at which the contact breaks under tension is given by $\xi_t = \frac{1}{2}\frac{1}{6^{1/3}}\frac{a^2}{R^*}$. For the JKR model, H{\ae}rvig et al. \cite{haervig2017adhesive} proposed a scaling law where the surface energy density is scaled with the modified effective Young's modulus as $\gamma = \gamma_r{(E/E_r)}^{2/5}$, where $\gamma_r  = 32$ mJ/m$^2$ is the real surface energy density of PA12 \cite{bahrami2022experimental,novak2006surface} and $E_r$ is the real Young's modulus of PA12.

 \subsection{Heat transfer for multi-sphere particles}
\begin{figure}[hbtp]
\centering
\includegraphics[width=0.6\textwidth]{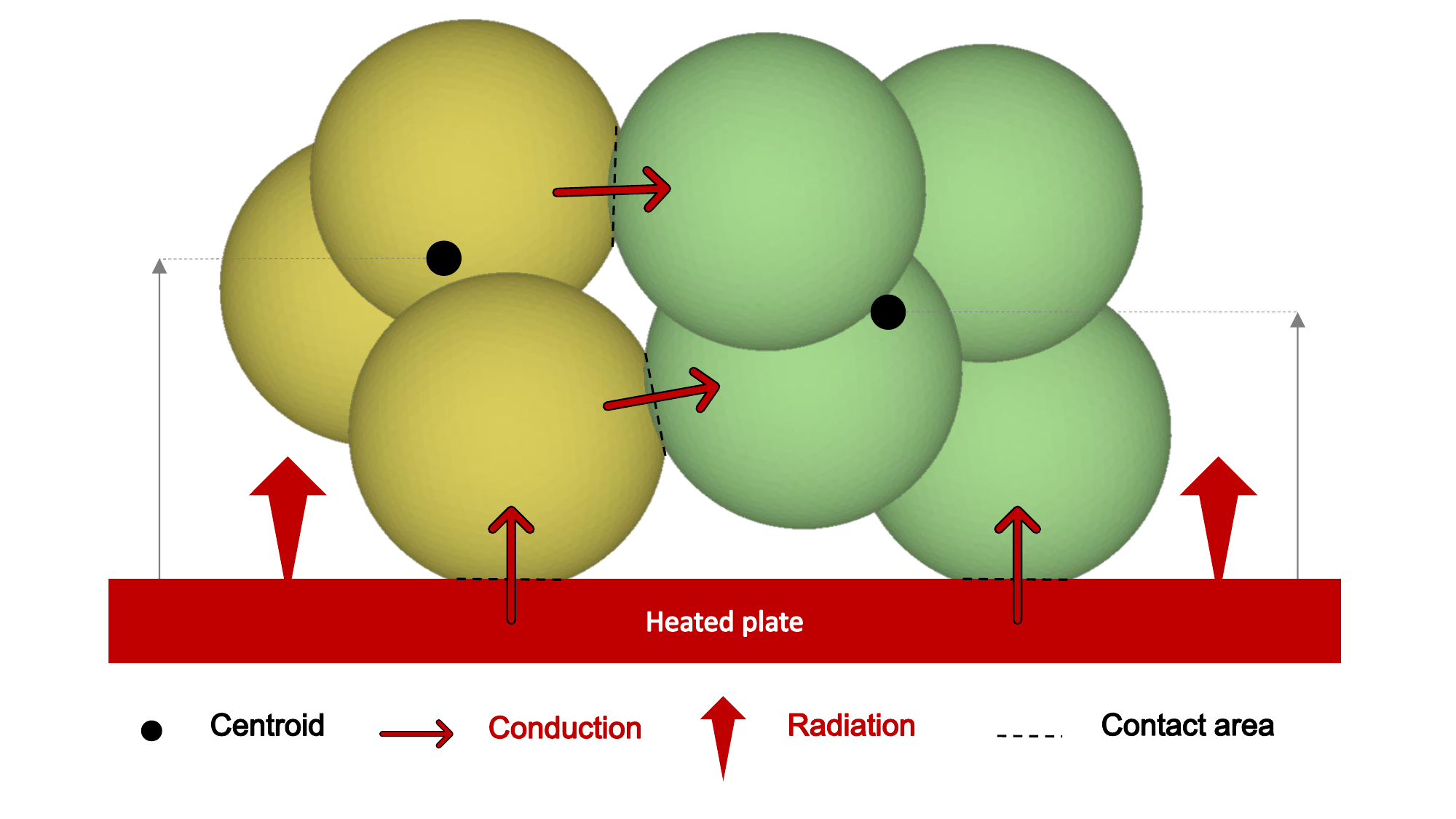}
\captionsetup{justification=justified}
\caption{Sketch for heat transfer mechanism of multi-spheres for conduction and radiation.}
\label{fig:heattransfer}
\end{figure}

The heat transfer for multi-sphere particles can include conduction, convection, and radiation. Conduction occurs when a clump overlaps with another clump or a boundary. While convection and radiation occur for the exposed portion of a clump's surface with the other particles and the ambient. Here, we consider the spreading of pre-heated powder over the surface of the printed part that has a temperature at the melting point. In this case, the temperature of the powder and the ambient is assumed to be the same and major heat exchange should occur between the particles and the solidified part. We do not consider thermal convection here as we do not expect strong upward convection current from spreading. For each clump $i$, a single temperature, $T_i$, is assigned, and is updated by

\begin{equation}
     \frac{\partial T_i}{\partial t}=\frac{Q_i}{m_ic_p},
\label{eq:temp}
\end{equation}

where $c_p$ is the heat capacity and $Q_i$ is the total thermal transfer rate.  which can be expressed as 
\begin{equation}
     Q_i=\sum_{j=1}^{N}Q_{\mathrm{cond},ij} + Q_{\mathrm{rad},i},
\label{eq:therml}
\end{equation}

\noindent where $Q_{\mathrm{cond},ij}$, and $Q_{\mathrm{rad},i}$ are the respective rate for conduction and radiation, which are demonstrated in \autoref{fig:heattransfer}. The conduction rate for clump $i$ needs to be summed over the $N$ neighbouring clump $j$ that $i$ is in contact with. Note that a pair of clumps can have multiple contacts which are essentially overlaps of their slave spheres. Thus, $Q_{\mathrm{cond},ij}$ is calculated as the summation of the conduction rate of the slave particles for clump $i$, which is

\begin{equation}
     Q_{\mathrm{cond},ij}=\sum_{s=1}^{N_s}\sum_{n=1}^{N_{\mathrm{neb},s}}k\frac{s_{sn}}{r}(T_j-T_i),
\label{eq:therml}
\end{equation}

\noindent where $N_s$ is the number of slaves of clump $i$, $N_{\mathrm{neb},s}$ is the number of contacts a slave sphere $s$ has with the slave spheres of $j$, $k$ is the thermal conductivity, and $s_{sn}$ is the overlap area of a pair of slave spheres in contact. The temperature gradient is calculated at the clump level, with $T_j$ being the temperature of clump $j$ and $r$ being the centroid-to-centroid distance between clumps $i$ and $j$. As the heat conduction flux is proportional to the contact area of two contacting slave particles, it is strongly affected by the fact that a smaller Young's modulus is used here, and so the value of the conduction coefficient has to be scaled. To this end, we correct the conduction coefficient as $k=k_r(E/E_r)^{{2}/{3}}$, where $k_r$ and $E_r$ are the realistic conductivity and modulus, and the ${2}/{3}$ exponent comes from the relation between contact area and the Young's modulus.

To derive this scaling, we assume two spherical PA12 particles of equal size in equilibrium, where one particle is resting on top of the other under its self weight. For the real Young’s modulus, we can calculate the deformation and contact area for these particles in contact via the Hertzian contact theory. For the (softer) system with scaled Young’s modulus, the contact area between the particles will be larger. Hence, since the thermal conduction is proportional to the contact area, we scale the thermal conduction coefficient of the system with softer Young's modulus, so that the thermal conduction of the scaled and the real system are the same, leading to the scaling law proposed above.

For radiation, we consider the heat exchange between particles and the hot substrate (solidified part from previous spreading cycle), and the the respective heat transfer rate can be approximated as 

\begin{align*}
&Q_{\mathrm{rad},i} = \epsilon\sigma_{SB}A_{ix}(T_i^4-T_\mathrm{substrate}^4),
\end{align*}

\noindent where $\epsilon$ is the material emissivity, $\sigma_{SB}=5.67\times10^{-8}$\,W/(m$^2/K^4$) is the Stefan-Boltzmann constant \cite{boltzmann1978ableitung}, and $T_\mathrm{substrate} = 443$ K represent the temperature of the substrate \cite{li2021experimental,laumer2015influence,peyre2015experimental}. The exposed area of each clump is approximated as $A_{ix}=G(z)A_i/2$, where $A_i$ is the surface area of clump $i$, and $G(z)$ is a height dependent shielding factor accounting for the blockage of radiation from the substrate to particles in the upper levels. The dividing factor of two comes from an approximation that assumes that on average the lower half of a clump is exposed. The functional form for $G(z)$ is given by $G(z) = 0.5\bigg(1-\mathrm{erf}\frac{z-D_{50}}{D_{50}/16}\bigg)$, where $D_{50} = 55 \,\mu$m is the mean diameter of the non-spherical particles and $D_{50}/16$ is the width of the error function which is considered here based on the requirement of a smooth transition of the radiation heat transfer across the bed in the $z$ direction. Inter-particle radiation is neglected on the assumption that the temperature gradient between particles is small. Admittedly, the real situation is more complicated, and the additional convection and radiation can be modelled by adapting existing algorithms for spherical particles, via switching in the surface area of the clumps~\citep{zhou2023cfd,cheng2013particle,wu2016effect}. 

\section{Effect of number of subspheres on heat conduction}

Representing an irregularly-shaped particle with different number of subspheres ($N_s$) can have an influence on its behavior, especially when it comes to the number of contacts with a neighboring particle or boundary and the effective stiffness of the multi-sphere, both of which influence heat conduction. Per our multi-sphere construction, as $N_s$ increases, the average size of subspheres decreases,  as every additional sphere is of smaller or equal size to the previous one, which in turn decreases the prefactor $\rho$ in \autoref{Eq:Hertz_rho} for the normal contact force defined in \autoref{Eq:Hertz_normal}. For a larger subsphere, a larger overlap area can be expected under the same load. On the other hand, having more subspheres potentially leads to a higher number of contacts, reducing the load on each subsphere. It is hard to derive the dependence of heat conduction (the total contact area) on $N_s$ through these two effects, as other factors also come to play, including the specific multi-sphere generation method as well as the shape and orientation of the particles at contact, or how densely they are packed.

\begin{figure*}[htb!]
\centering
\begin{subfigure}{0.48\linewidth}
    \includegraphics[trim={0cm 0cm 0cm 0cm},clip,width=\linewidth]{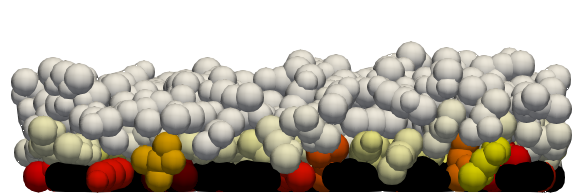}
    \subcaption{}
\end{subfigure}
\hfill
\begin{subfigure}{0.48\linewidth}
    \includegraphics[trim={0cm 0cm 0cm 0cm},clip,width=\linewidth]{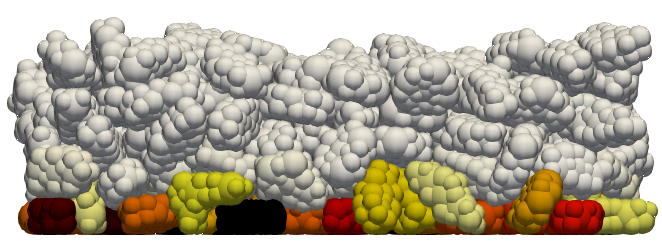}
    \subcaption{}
\end{subfigure}
\begin{subfigure}{0.1\linewidth}
    \includegraphics[trim={0cm 0cm 0cm 0cm},clip,width=\linewidth]{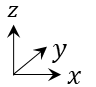}
\end{subfigure}
\hspace{10pt}
\begin{subfigure}{0.48\linewidth}
    \includegraphics[trim={0cm 0cm 0cm 0cm},clip,width=\linewidth]{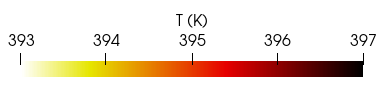}
\end{subfigure}
\caption{Snapshot after $0.50$ s of simulation of a sample layer of multi-spheres with (a) $5$ subspheres and (b) $25$ subspheres. The particles are coloured according to their temperature $T$. The simulations are conducted for $\alpha = 2$.}
\label{fig:subspheres_graphic} 
\end{figure*}
Alternatively, we perform auxiliary simulations of a sample layer of multipshere particles in a sample box with a hot substrate and periodic boundary conditions in the lateral directions. Radiation is not included in this test. For each test, we generate loose packings of particles with a certain $N_s$, settle under gravity on the hot substrate and monitor the temperature rise due to heat conduction with the hot substrate. The initial particle temperature is $T_\mathrm{in}=393$\,K and the wall temperature is fixed to $443$ K. Two examples using $N_s=$5 and $N_s=$25 are shown in \autoref{fig:subspheres_graphic}. The particles are colored by their temperature, which shows that the bottom layer of particles are heated up by the hot boundary. The temperature difference between the bottom particle layer and the upper layers is still much smaller than the difference with the hot wall, which explains the lack of temperature rise in the upper layers. In both cases, some variations in temperature can be seen in the bottom layer, with the more horizontally orientated particles being heated up more significantly due to the increased number of contacts with the bottom in comparison to the more vertically orientated particles. In addition, packing of particles with more subspheres is less dense as reflected by the increased layer height, which could be a result of the change of effective particle surface roughness and cohesion, which is also a function of subsphere size.

\begin{figure*}[htb!]
\centering
\begin{subfigure}{0.48\linewidth}
    \includegraphics[trim={0cm 0cm 0cm 0cm},clip,width=\linewidth]{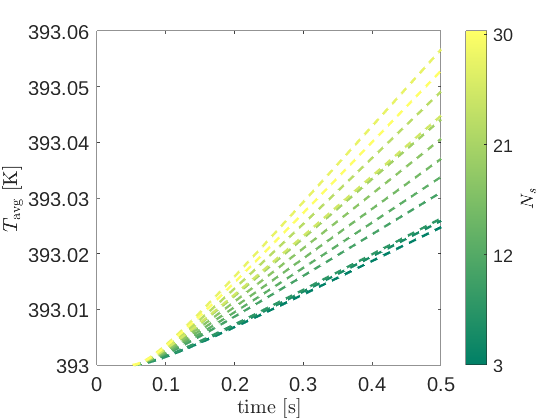}
    \subcaption{}
\end{subfigure}
\hfill
\begin{subfigure}{0.48\linewidth}
    \includegraphics[trim={0cm 0cm 0cm 0cm},clip,width=\linewidth]{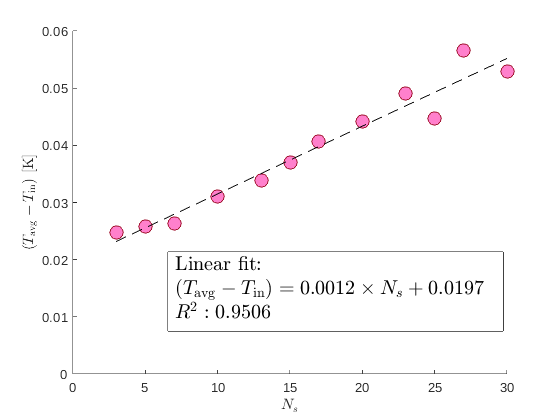}
    \subcaption{}
\end{subfigure}
\caption{The rise of average temperature of a sample particle layer with conduction only for particles made of different numbers of subspheres. The dashed line indicates a linear regression fitting line $(T_\mathrm{avg}-T_\mathrm{in}) = 0.0012\times{N_s} + 0.0197$ as a potential correction method.}
\label{fig:subsphere_plot} 
\end{figure*}
To quantify the influence of $N_s$, we plot the average particle temperature in the layer, $T_{ave}$, vs. time in \autoref{fig:subsphere_plot}(a), with the curves colored by the number of subspheres. Heat conduction begins at about $t=0.05$\,s after the particles settle down, and we focus on the initial period up to $t=0.50$\,s, during which the temperature rise is significantly less than the $50$\,K temperature difference between the particles and the hot substrate, and the temperature rise is therefore linear in time. The result clearly indicate that heat conduction is more significant for larger $N_s$, and we quantify the temperature rise, $T_\mathrm{ave}-T_\mathrm{in}$, at $t=0.50$\,s for different  as shown in \autoref{fig:subsphere_plot}(b). The temperature rise, which is roughly proportional to the total contact area between the particles and the substrate, doubles as $N_s$ increases from 3 to 30. This means that heat conduction is artificially enhanced by using a larger $N_s$ as a result of subsphere softening and increased number of contact, despite having looser packing. As the dependence of the temperature rise is rather linear with $N_s$, it is possible to empirically scale the conductivity to remove the dependence of $N_s$ according to the fitting in \autoref{fig:subsphere_plot}(b). However, more rigorous solutions potentially exist with controlling the subsphere size during multi-sphere generation, which is beyond the scope of current study. In the rest of the paper, we use results of $N_s=10$. The thermal radiation, on the other hand, is much less sensitive to $N_s$ as it is not a function of contact area.

\section{Simulation of powder spreading}
\subsection{Numerical setup}

\begin{figure*}[htb!]
\centering
\begin{subfigure}{\linewidth}
    \includegraphics[trim={0cm 0cm 0cm 0cm},clip,width=0.95\linewidth]{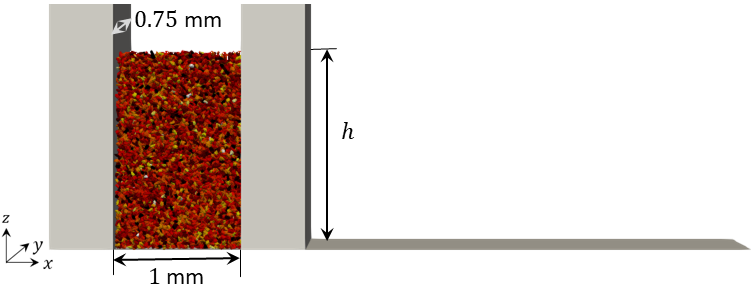}
    \subcaption{}
\end{subfigure}
\begin{subfigure}{\linewidth}
    \includegraphics[trim={0cm 0cm 0cm 0cm},clip,width=\linewidth]{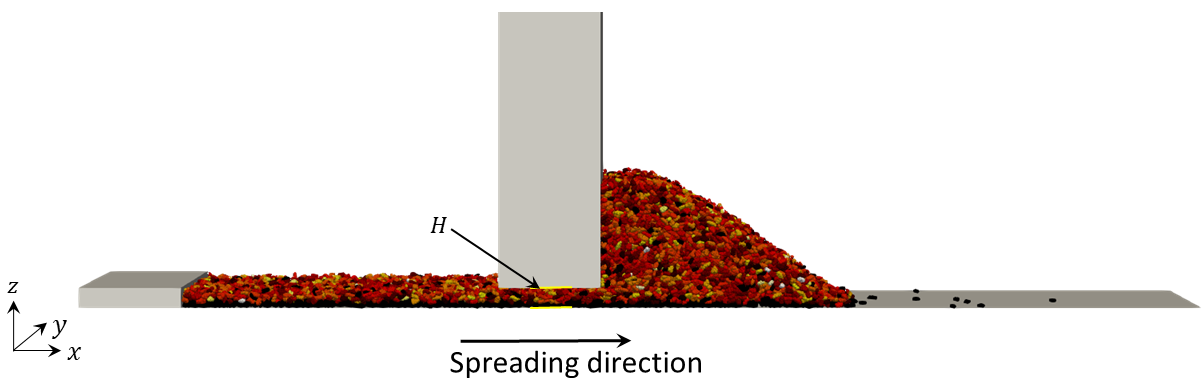}
    \subcaption{}
\end{subfigure}
\hfill
\begin{subfigure}{\linewidth}
    \includegraphics[trim={0cm 0cm 0cm 0cm},clip,width=\linewidth]{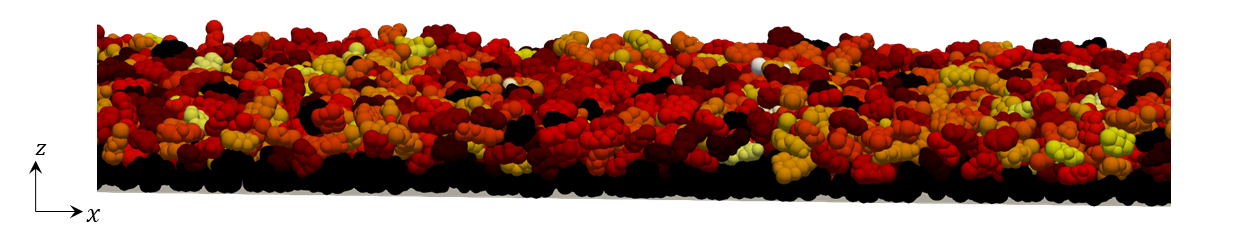}
    \subcaption{}
\end{subfigure}
\begin{subfigure}{\linewidth}
    \includegraphics[trim={0cm 0cm 0cm 0cm},clip,width=\linewidth]{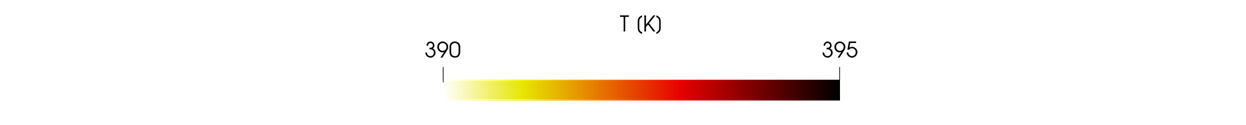}
\end{subfigure}
\caption{Powder spreading setup showing (a) side view during initial particle deposition, (b) side view during spreading and (c) deposited powder layer structure after $0.05$ s of powder spreading process with a blade velocity $v = 50$ mm/s.}
\label{fig:powderSpreading}
\end{figure*}
A thin layer of the powder bed of width $0.75$ mm is simulated using periodic boundary conditions in the $y$-direction. In this numerical set-up, the substrate is assumed to be flat and frictional. We inserted ca. $20000$  multi-sphere particles with a mean diameter $D_{50} = 55 ~\mathrm{\mu}$m as shown in \autoref{fig:powderSpreading}(a) in front of the spreader tool, at $(x,y,z) \in [0.5,1.5]$ mm $\times [0,0.75]$ mm $\times [0,h]$ mm until the total bulk particle volume equals $0.7$ mm$^3$, which is sufficient to create a powder layer of $5$ mm length, $0.75$ mm width and $0.1$ mm height. In the beginning of the simulations, the particles are allowed to deposit and relax under the effect of gravity. Then, the spreading process begins with the tool starting to move the at a constant speed $v$. The particles are spread in a layer where the tool gap is set to $H = 100 \, \mu$m, which corresponds to about $1.8D_{50}$ in $z-$direction as shown in \autoref{fig:powderSpreading}(b). The structure of the deposited powder layer is shown in \autoref{fig:powderSpreading}(c). Particles reaching the end of the powder bed (at $x = 5$ mm) are not used for the analysis. The powder layer, spread on the hot substrate, undergoes heating through both conduction and radiation. Finally, the simulation is stopped when the temperature of the spread powder layer stabilizes, indicating minimal temperature changes. Thus, the maximum simulation time is chosen as  $t_\mathrm{max} = 1$ s, taking into account the time for initial settling and relaxation of particles, the spreading time and the time for the temperature of the powder layer to achieve a stabilized state.
\subsection{Parameters for powder spreading simulations}
The packing density of the powder layers decreases with increasing spreading velocity, as reported by Nasato et al. \cite{nasato2020influence}. Consequently, the effective thermal conductivity of the powder bed at the bulk scale should vary with the spreading velocity. Therefore, we analyze the effect of spreading blade velocity on the temperature evolution of the deposited powder layer, considering two blade velocities: $v = 50$ and $250$ mm/s.

Above the glass transition temperature of $322$ K (i.e. $48.8^\circ$C), the Young's modulus of PA12 undergoes a sharp decline as a function of temperature \cite{bai2014influence}. This temperature-induced decrease in the Young's modulus is modelled as a function of temperature by \autoref{eq:functionModulus}, where $\alpha = 2$ is a commonly used model as proposed by Luding et al. \cite{luding2005discrete}, however little experimental evidence exists to verify this value. To investigate the effect of the Young's modulus decrease on the powder layer's local structure, we conducted simulations for various $\alpha$ values: $2, 20,$ and $200$.

Particle shapes play a crucial role in the packing density as well as in the contact area between the particles and between the particles and the substrate. Non-spherical particles of PA12 have a higher contact area with the substrate as well as inter-particle contact area, leading to a higher conductivity of the powder layer compared to spherical particles. Spherical particles of PA12, on the other hand, are often used in the powder spreading process due to their better flowability. Therefore, we analyze the effect of particle shapes on the temperature evolution of the deposited powder layer, considering real PA12 particle shapes and spherical particles of the same volumes of PA12.
\subsection{Temporal evolution of powder bed temperature}
In this sub-section, we examine the temporal evolution of average temperature $T_\mathrm{avg}$ within a specific section of the powder layer ($1$ to $2$ mm in length) under various parametric conditions during powder spreading. From the temporal evolution trend, we assess the temperature attained after $1.0$\,s of powder spreading simulation in different parametric scenarios.
\subsubsection{Effects of blade velocity}
\begin{figure*}[htb!]
\centering
\begin{subfigure}{0.48\linewidth}
    \includegraphics[trim={0cm 0cm 0cm 0cm},clip,width=\linewidth]{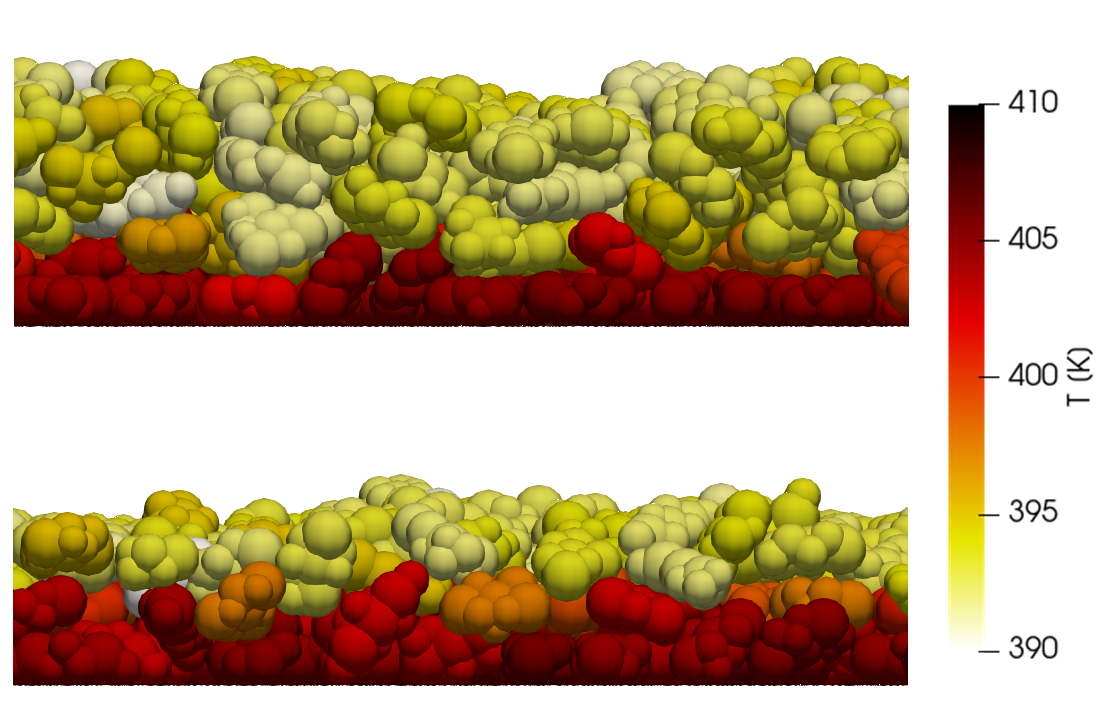}
    \subcaption{}
\end{subfigure}
\hfill
\begin{subfigure}{0.48\linewidth}
    \includegraphics[trim={0cm 0cm 0cm 0cm},clip,width=\linewidth]{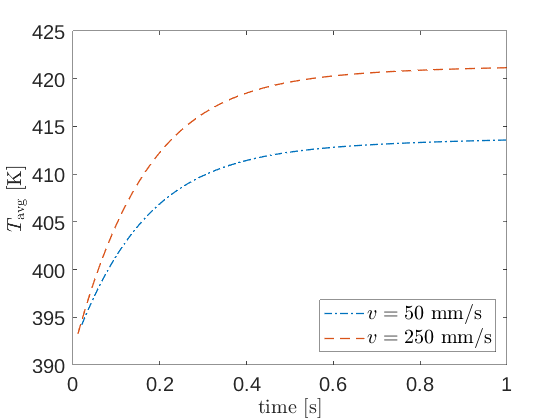}
    \subcaption{}
\end{subfigure}
\caption{(a) Packing of powder layers for different blade velocities $v = 50$ mm/s (top) and $v = 250$ mm/s (bottom) and (b) average temperature of the powder bed $T_\mathrm{avg}$ as a function of time for simulations with the two different blade velocities.}
\label{fig:Velocity} 
\end{figure*}

In \autoref{fig:Velocity}(a), the packing structure of a deposited powder layer is visualized at two distinct blade velocities with uniform resolution. Notably, the powder layer at a blade velocity of $250$\,mm/s appears thinner compared to the layer at $50$\,mm/s. Simultaneously, in \autoref{fig:Velocity}(b), the mean temperature evolution within a section of the powder layer is shown over a time period of $1.0$\,s. The powder layer subjected to a blade velocity of $250$\,mm/s exhibits a notably higher mean temperature compared to the layer at $50$\,mm/s. 
The faster heating rate and the higher temperature for higher blade velocity are primarily due to two reasons. First, as powders are being spread faster, more particles are brought to contact with the substrate given the same amount of time, thus $T_{avg}$ rises faster. Second, the thinner layer with the faster spreading means that heat can propagate through the height of the layer faster. The thinner layer also results in a higher mean temperature $T_{avg}$ of the powder bed.

\subsubsection{Effects of degree of Young's modulus change}
\begin{figure*}[htb!]
\centering
\begin{subfigure}{0.48\linewidth}
    \includegraphics[trim={0cm 0cm 0cm 0cm},clip,width=\linewidth]{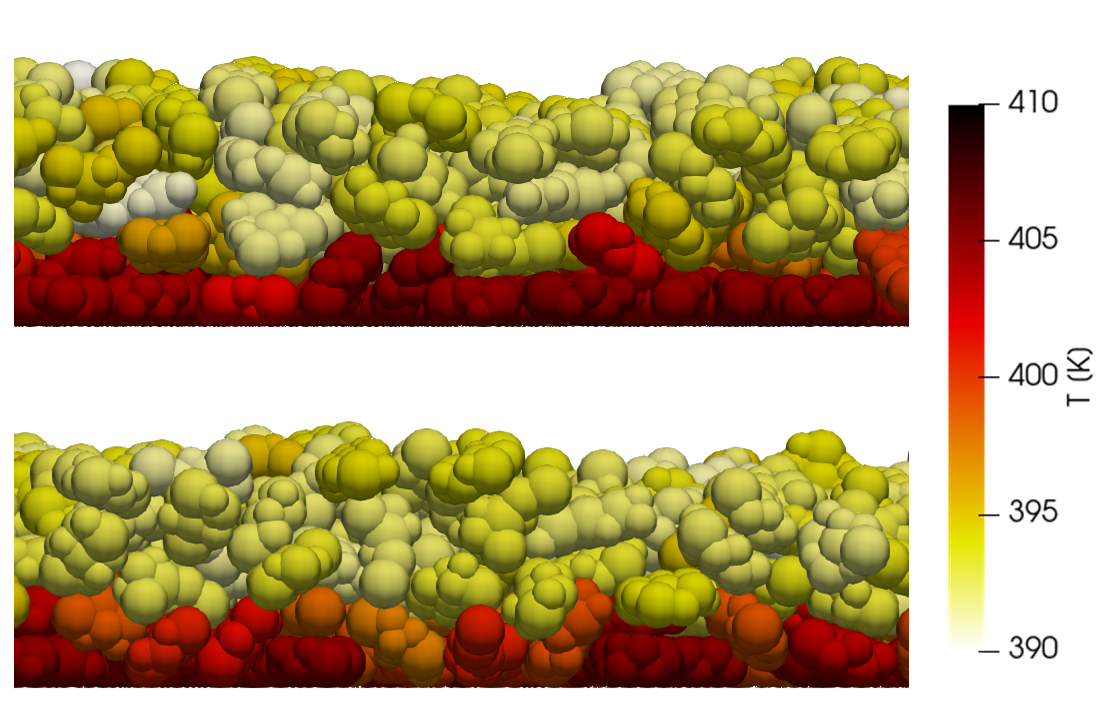}
    \subcaption{}
\end{subfigure}
\hfill
\begin{subfigure}{0.48\linewidth}
    \includegraphics[trim={0cm 0cm 0cm 0cm},clip,width=\linewidth]{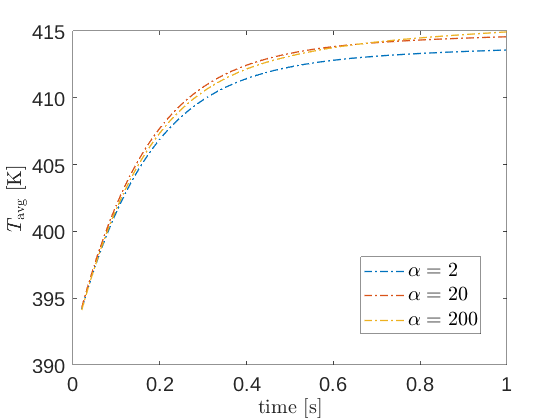}
    \subcaption{}
\end{subfigure}
\caption{Packing of powder layers for different  degree of Young's modulus change parameter $\alpha = 2$ (top) and $\alpha = 200$ (bottom) and (b) average temperature of the powder bed $T_\mathrm{avg}$ as a function of time for simulations with different $\alpha$.}
\label{fig:alpha} 
\end{figure*}

In \autoref{fig:alpha}(a), the packing structure of the deposited powder layer is depicted for two degree of Young's modulus change parameter, denoted by $\alpha$.  Interestingly, the powder layer structures remain notably similar for $\alpha = 2$ and $\alpha = 200$. This visual similarity is further confirmed by the analysis of the mean temperature evolution illustrated in \autoref{fig:alpha}(b) over a time frame of $1.0$ s. Here, we observe the evolving temperature is slightly lower for $\alpha = 2$, but no substantial differences in the temperature profiles are observed for $\alpha = 20$ and $200$ values. The higher temperatures observed for $\alpha = 20$ and $200$ are primarily attributed to increased particle softening with rising temperatures. This phenomenon results in a greater overlap of particles, leading to enhanced heat conduction between the substrate and the powder bed.

\subsubsection{Effects of particle shapes}
\begin{figure*}[htb!]
\centering
\begin{subfigure}{0.48\linewidth}
    \includegraphics[trim={0cm 0cm 0cm 0cm},clip,width=\linewidth]{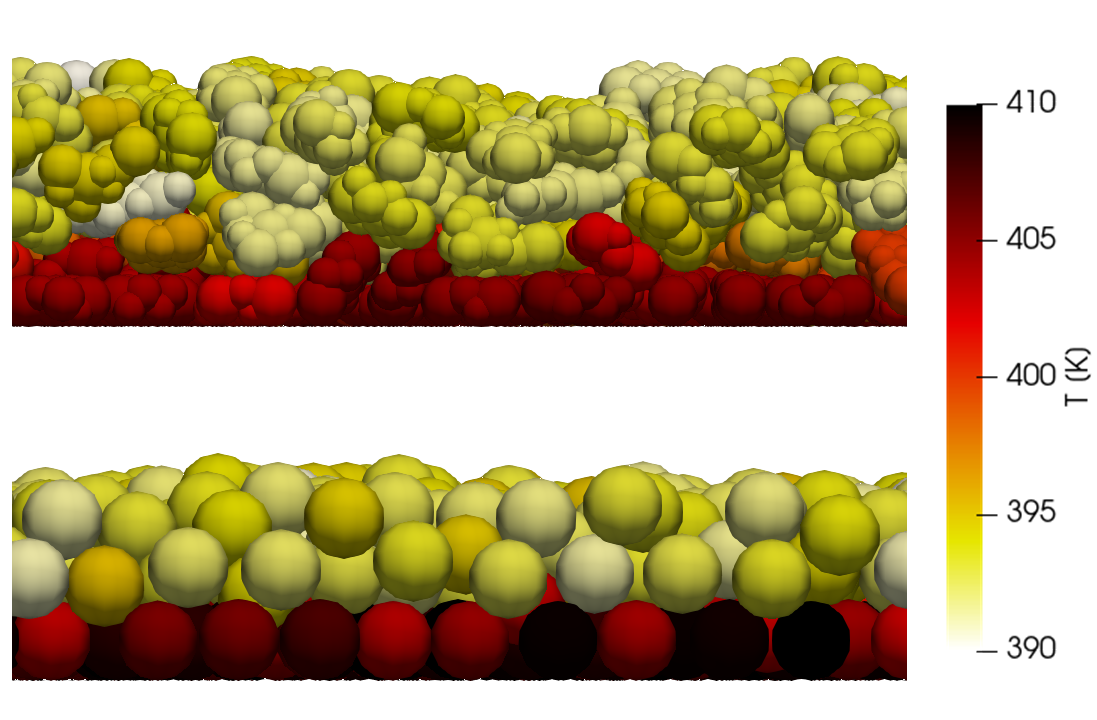}
    \subcaption{}
\end{subfigure}
\hfill
\begin{subfigure}{0.48\linewidth}
    \includegraphics[trim={0cm 0cm 0cm 0cm},clip,width=\linewidth]{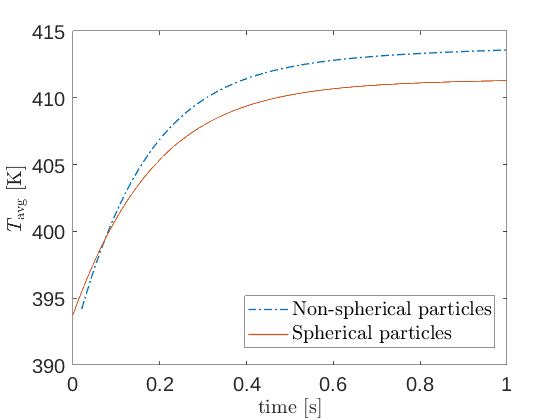}
    \subcaption{}
\end{subfigure}
\caption{Packing of powder layers for different particle shapes: non-spherical particles (top) and spherical particles (bottom) and (b) average temperature of the powder bed $T_\mathrm{avg}$ as a function of time for simulations with the two different particle shapes with $v = 50$ mm/s and $\alpha = 2$.}
\label{fig:sphere} 
\end{figure*}

In \autoref{fig:sphere}(a), the packing structure of a deposited powder layer is presented, showcasing two distinct particle shapes: irregular shaped particles and regular spherical particles. Notably, the packing structure involving irregular particles displays a significantly larger contact area, both among the particles themselves and with the heated substrate. This structural difference is further explored in \autoref{fig:sphere}(b), which illustrates the evolution of the mean temperature within the identical section of the powder layer over a duration of $1.0$ s. As expected, owing to the larger contact area, the powder layer comprising irregular particles exhibits a notably higher mean temperature in contrast to the layer composed of spherical particles.
\subsection{Spatial evolution of powder bed temperature}
In this sub-section, we explore the in-situ average temperature of particles beneath the blade $T_\mathrm{blade}$ during the powder spreading process.
\begin{figure*}[htb!]
\centering
\begin{subfigure}{0.48\linewidth}
    \includegraphics[trim={0cm 0cm 1.2cm 0cm},clip,width=\linewidth]{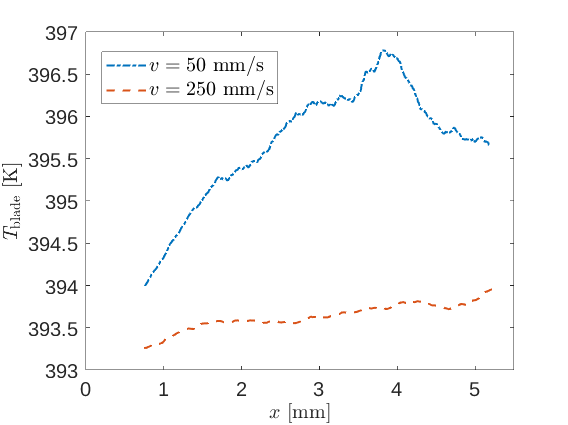}
    \subcaption{}
\end{subfigure}
\hfill
\begin{subfigure}{0.48\linewidth}
    \includegraphics[trim={0cm 0cm 1.2cm 0cm},clip,width=\linewidth]{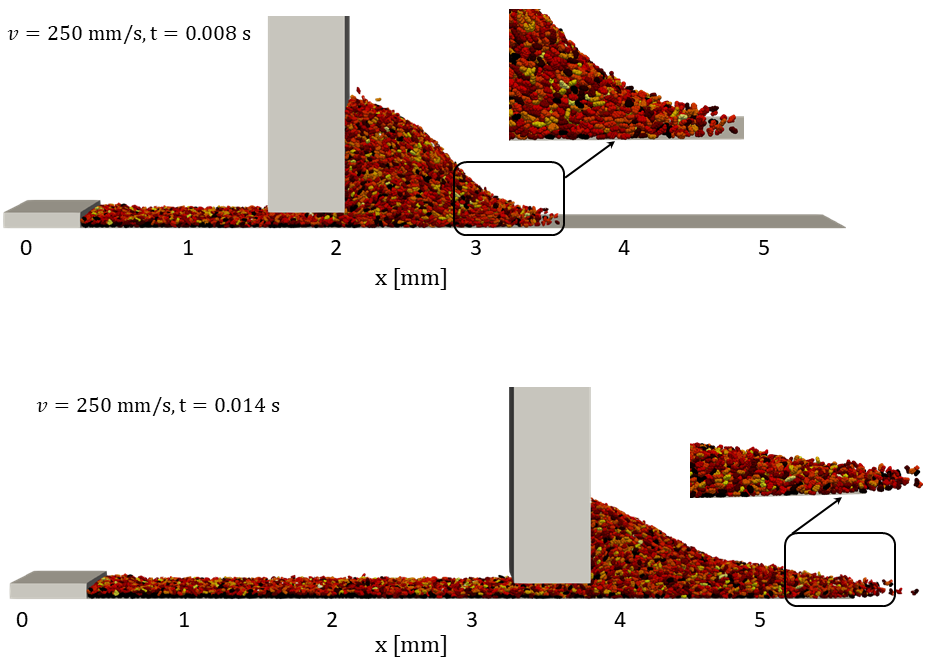}
    \subcaption{}
\end{subfigure}
\caption{(a) Temperature of powder layer under the moving blade as a function of the spreading length for two different blade velocities $50$ mm/s and $250$ mm/s. (b) Powder spreading after $t =$ s and $t = $ s for blade velocity $250$ mm/s.}
\label{fig:VelocityTemperatureSpace} 
\end{figure*}
\autoref{fig:VelocityTemperatureSpace}(a) illustrates the comparison of the average particle temperature beneath the moving blade in relation to the spreading length for two distinct blade velocities. At a blade velocity of $250$ mm/s, the faster motion restricts the duration for particles to absorb heat, resulting in a notably lower temperature profile in contrast to the $50$ mm/s blade velocity. The temperature of the particles demonstrates a consistent increase concerning the spreading length $x$. Conversely, the temperature profile exhibits an initial steady rise for the blade velocity of $50$ mm/s, followed by a sharp decline in temperature for $x > 4$ mm. This decline can be attributed to the movement of cooler particles within the heap in front of the blade. In cases of a substantial heap, cooler particles from the heap's surface cascade downward, settling well ahead of the blade. This provides ample time for heat absorption. As the powder spreading process nears completion and the heap diminishes, cooler particles settle closer to the blade, reducing the time available for heat absorption and resulting in a corresponding temperature decrease. Higher blade velocity leads to a faster spreading process, offering minimal time for heat absorption. \autoref{fig:VelocityTemperatureSpace}(b) illustrates the flow profile of particles corresponding to the higher blade velocity.
\begin{figure*}[htb!]
\centering
\begin{subfigure}{0.48\linewidth}
    \includegraphics[trim={0cm 0cm 1.2cm 0cm},clip,width=\linewidth]{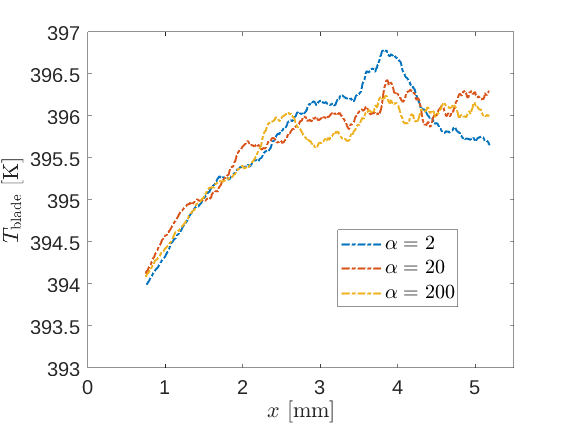}
\end{subfigure}
\caption{Temperature of powder layer under the moving blade as a function of the spreading length for different $\alpha$ with spreading velocity $50$ mm/s for non-spherical particles.}
\label{fig:AlphaTemperatureSpace} 
\end{figure*}

\begin{figure*}[htb!]
\centering
\begin{subfigure}{0.48\linewidth}
    \includegraphics[trim={0cm 0cm 1.2cm 0cm},clip,width=\linewidth]{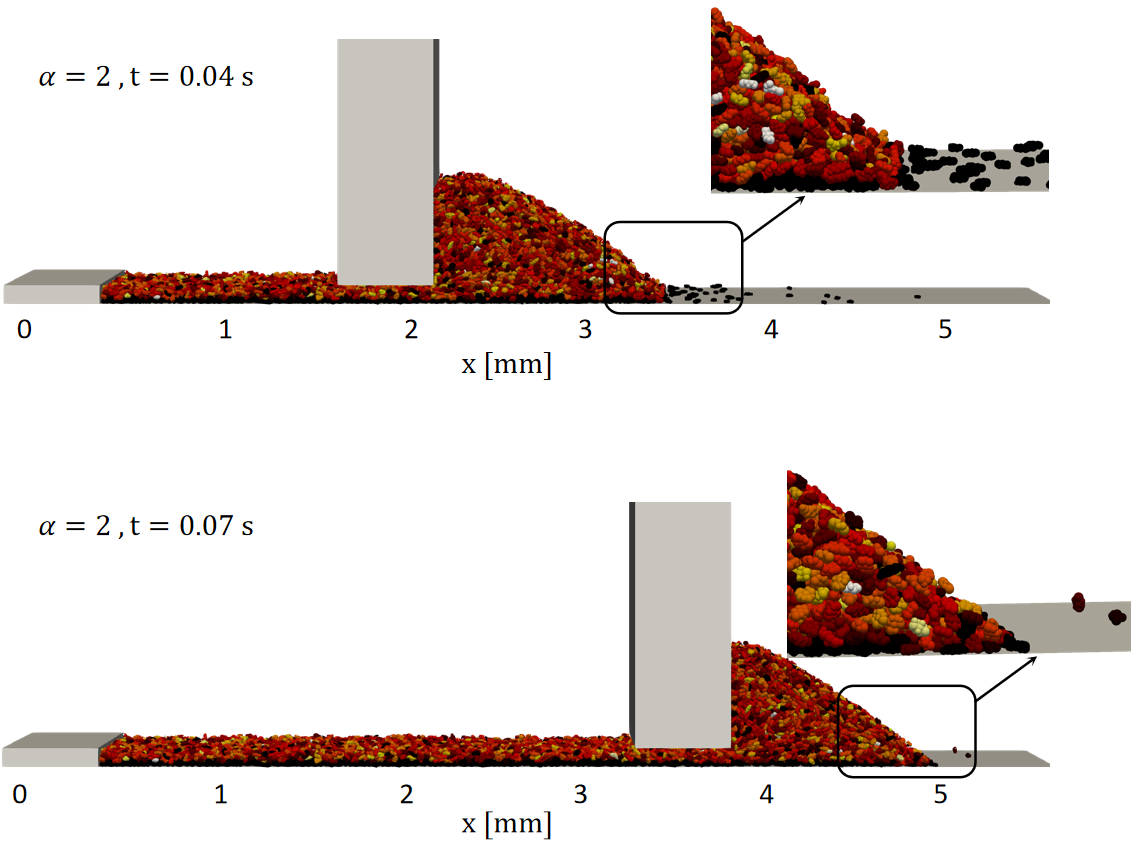}
    \subcaption{}
\end{subfigure}
\hfill
\begin{subfigure}{0.48\linewidth}
    \includegraphics[trim={0cm 0cm 1.2cm 0cm},clip,width=\linewidth]{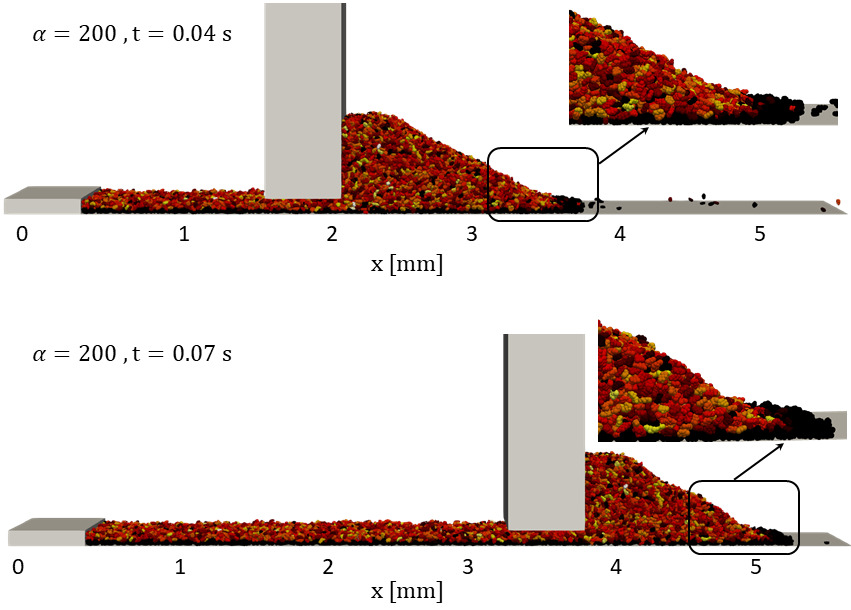}
    \subcaption{}
\end{subfigure}
\hfill
\begin{subfigure}{0.8\linewidth}
    \includegraphics[trim={0cm 0cm 1.2cm 0cm},clip,width=\linewidth]{figures/PowderSpreadingGranularMatterRevisedBotttom.png}
\end{subfigure}
\caption{Powder spreading after $t = 0.04$ s and $t = 0.07$ s for (a) $\alpha = 2$ and (b) $\alpha = 200$.}
\label{fig:AlphaSchematic} 
\end{figure*}

In \autoref{fig:AlphaTemperatureSpace}, it can be observed that the temperature profile exhibits an initial steady rise for $\alpha = 2$, followed by a sharp decline in temperature for $x > 4$ mm as also observed in \autoref{fig:VelocityTemperatureSpace}(a). In contrast, the temperature profile increases with $x$ for $\alpha = 20$ and $200$ and there are no significant drop in temperature for $x > 4$ mm for the two cases. This suggests that the temperature sensitivity of Young's modulus play a critical role in affecting heat transfer during the in-situ powder spreading process. Further, we visualise the local temperature of particles during the spreading process as shown in \autoref{fig:AlphaSchematic}(a) and (b) for the two extreme cases of $\alpha = 2$ and $\alpha = 200$. For $\alpha = 2$, the hot particles at the tip of the heap get overrun by the intermittent avalanche of cold particles from the top of the heap. In cases of a substantial heap, cooler particles from the heap's surface cascade downward, settling well ahead of the blade. This provides ample time for heat absorption. As the powder spreading process nears completion and the heap diminishes, cooler particles settle closer to the blade, reducing the time available for heat absorption and resulting in a corresponding temperature decrease. Therefore, the temperature of the particles at the tip of the heap is relatively lower as observed in \autoref{fig:AlphaSchematic}(a). For the higher alpha case, the flow is more cohesive and the heap of particles in front of the blade remains quite rigid. Therefore, the temperature of the particles at the tip of the heap is consistently higher as observed in \autoref{fig:AlphaSchematic}(b). Therefore, the temperature profile of the particles during spreading consistently rises with spreading distance $x$.

\begin{figure*}[htb!]
\centering
\begin{subfigure}{0.48\linewidth}
    \includegraphics[trim={0cm 0cm 1.2cm 0cm},clip,width=\linewidth]{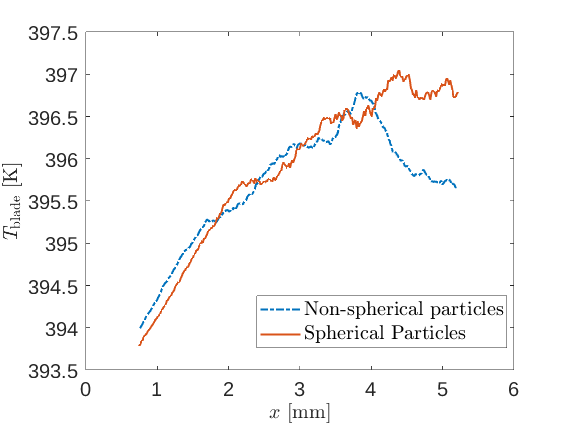}
     \subcaption{}   
\end{subfigure}
\hfill
\begin{subfigure}{0.48\linewidth}
    \includegraphics[trim={0cm 0cm 1.2cm 0cm},clip,width=\linewidth]{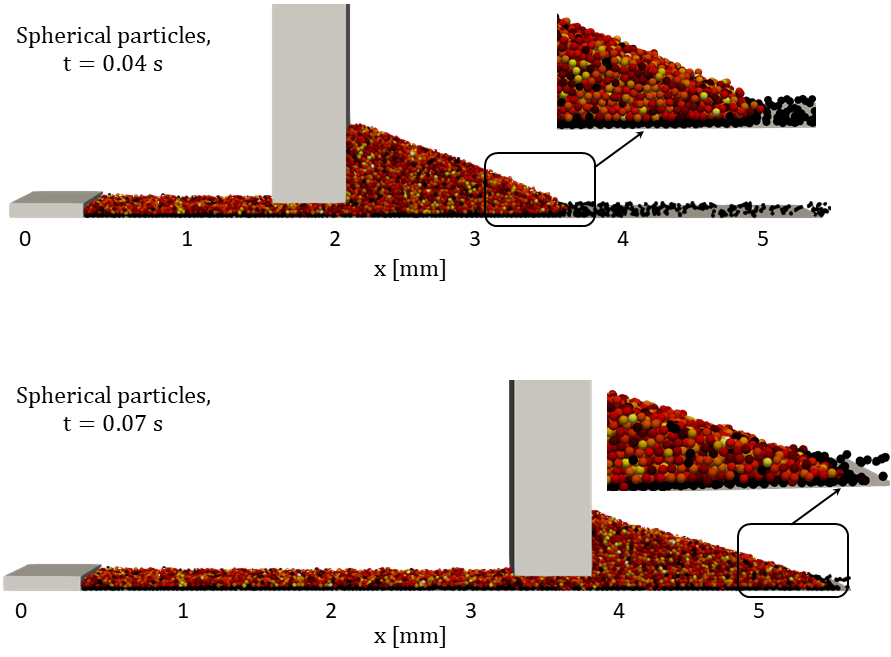}
    \subcaption{}
\end{subfigure}
\hfill
\caption{(a) Temperature of powder layer under the moving blade as a function of the spreading length for particles with spherical and irregular shapes. (b) Powder spreading after $t = 0.04$ s and $t = 0.07$ s for spherical particles.}
\label{fig:ShapeTemperatureSpace} 
\end{figure*}

Next, we show the average particle temperature beneath the moving blade $T_\mathrm{blade}$ compared for non-spherical and spherical particles for a blade velocity $50$ mm/s and $\alpha = 2$ as shown in \autoref{fig:ShapeTemperatureSpace}(a). Interestingly, although for both cases $\alpha = 2$, the temperature profile of the spherical particles do not show in drop for $x > 4$ mm. This is because, spherical particles have higher flowability and hence cold particles from the top of the heap get deposited much farther ahead of the tip of the heap. As a result, the tip of the heap consistently have particles at higher temperature, resulting in a consistent increase in $T_\mathrm{blade}$ as a function of $x$. \autoref{fig:ShapeTemperatureSpace}(b) illustrates the flow profile of spherical particles, which can be compared to that of irregularly shaped particles in \autoref{fig:AlphaSchematic}(a). Clearly, the cold spherical particles from the top of the heap get deposited much further ahead of the tip of the heap compared to the deposition of irregularly shaped particles.

\section{Conclusions}
In this study, we enhanced the multi-sphere algorithm by integrating it with a thermal particle formulation within the open-source code \texttt{MercuryDPM}. This modification allowed us to comprehensively explore the interplay between heat transfer and properties of irregularly-shaped powders in the spreading process. Utilizing the refined algorithm, we systematically examined the impact of blade speed, particle stiffness, and particle shapes on the temperature profiles of the powder bed.

For the temporal evolution of the average temperature, $T_\mathrm{avg}$, within a specific section of the spread powder layer, higher blade velocity leads to faster heating rates during powder spreading and higher temperatures of the powder layer due to increased particle-substrate contact and faster heat propagation through the thinner powder layer. As for the temperature dependence of the material modulus, a marginal decrease in temperature is noted for $\alpha = 2$, while no significant variations in temperature profiles are observed for $\alpha = 20$ and $200$. Higher temperatures for the latter cases are attributed to increased particle softening, fostering greater particle overlap and improved heat conduction between the substrate and the powder bed. For the particle shape, the layer with irregular particles exhibits a significantly higher evolving temperature compared to the layer with spherical particles, attributed to the larger contact area of irregular particles.

We explored the in-situ temperature of particles beneath the blade, $T_\mathrm{blade}$, in relation to the spreading distance $x$ during the powder spreading process. 
For faster blade velocity, the temperature increases with spreading length as it leads to a wider spreading. For slower blade velocity ($\alpha=2$, irregular shape), the temperature initially rises with spreading length $x$ and then sharply declines beyond $4$ mm due to cooler particles from the diminishing heap. For different $\alpha$, e.g. $\alpha = 20$ and $200$, the temperature again consistently increases with spreading length, indicating that the sensitivity of Young's modulus significantly affects heat transfer during spreading. The local temperature dynamics depict distinct behaviors, with $\alpha = 2$ showing intermittent avalanches of cold particles impacting heat absorption, and $\alpha = 200$ displaying a more cohesive flow, resulting in consistently higher temperatures at the tip of the heap. For the comparison of different particle shape, we observe a lack of temperature drop for spherical particles beyond $4$ mm. This can be attributed to their higher flowability that leads to the deposition of colder particles further ahead of the tip of the heap to be heated up, ensuring a consistent increase in $T_\mathrm{blade}$ with spreading distance $x$.

\section*{Acknowledgement}
We would like to express our gratitude to the Deutsche Forschungsgemeinschaft (DFG, German Research Foundation) for their generous funding of the Collaborative Research Center 814 (CRC 814), Project Number 61375930-SFB 814 `Additive Manufacturing', sub-project B1. Additionally, we extend our thanks to the Humboldt Research Foundation for awarding the `Humboldt Research Fellowship'. This work received support from various institutions, including the Interdisciplinary Center for Nanostructured Films (IZNF), the Competence Unit for Scientific Computing (CSC), and the Interdisciplinary Center for Functional Particle Systems (FPS) at Friedrich-Alexander-Universität Erlangen-Nürnberg.

\section*{Conflicts of interests}
There are no conflicts to declare.







\bibliography{references}

\end{document}